\documentclass[twocolumn,showpacs]{revtex4}
\usepackage{amsmath}
\usepackage{times}
\usepackage{epsfig}
\usepackage{epstopdf}
\usepackage{graphicx}

\begin{document}

\draft

\title{Information filtering via biased heat conduction}

\author{Jian-Guo Liu}\thanks{liujg004@ustc.edu.cn}
\address{Research Center of Complex Systems Science, University of
Shanghai for Science and Technology, \\ Shanghai 200093, Peoples's
Republic of China} \affiliation{CABDyN Complexity Center, Sa\"{i}d
Business School, University of Oxford, Park End Street, Oxford, OX1
1HP, United Kingdom}

\author{Tao Zhou}\thanks{zhutou@ustc.edu}
\affiliation{Web Sciences Center, University of Electronic Science
and Technology of China, Chengdu 610054, Peoples's Republic of
China}


\author{Qiang Guo}
\affiliation{Research Center of Complex Systems Science, University
of Shanghai for Science and Technology, \\ Shanghai 200093,
Peoples's Republic of China}

\begin{abstract} \textnormal{\small {Heat conduction process has
recently found its application in personalized recommendation [T.
Zhou \emph{et al.}, PNAS 107, 4511 (2010)], which is of high
diversity but low accuracy. By decreasing the temperatures of
small-degree objects, we present an improved algorithm, called
biased heat conduction (BHC), which could simultaneously enhance the
accuracy and diversity. Extensive experimental analyses demonstrate
that the accuracy on MovieLens, Netflix and Delicious datasets could
be improved by 43.5\%, 55.4\% and 19.2\% compared with the standard
heat conduction algorithm, and the diversity is also increased or
approximately unchanged. Further statistical analyses suggest that
the present algorithm could simultaneously identify users'
mainstream and special tastes, resulting in better performance than
the standard heat conduction algorithm. This work provides a
creditable way for highly efficient information filtering.}}
\end{abstract}


\maketitle

With the advent of the Internet \cite{Broder2000} and wide
application of Web 2.0 techniques, there sprout many web sites that
enable large communities to aggregate and interact. For example,
Twitter allows its $1.7\times 10^8$ members to share interests and
life experiences, Facebook has already exceeded 500 million members
since July 16th, 2010, and their members are growing ever faster.
This brings massive amount of accessible information, more than
every individual's ability to process. Searching, filtering and
recommending thus become indispensable in the Internet era, in which
the personalized recommender systems have become an effective tool
to address the information overload problem by predicting users'
interests and habits based on their historical records. Personalized
recommender systems have been used to recommend books and CDs at
Amazon.com, movies at Netflix.com, and news at Versifi Technologies
(formerly AdaptiveInfo.com) \cite{Adomavicius2005}. Motivated by the
practical significance to e-commerce, recommender systems have
caught increasing attention and become an essential issue
\cite{Ecommerce2001,PNAS}. A personalized recommender system
includes three parts: data collection, model analysis and
recommender algorithm, where the algorithm is the core part. Thus
far, various kinds of algorithms have been proposed, including
collaborative filtering (CF) approaches
\cite{Herlocker2004,Konstan1997,Liu2009,Liu2008b,Dun2009,Liu2009int},
content-based analyses \cite{Balab97,Pazzani99}, tag-aware
algorithms \cite{Zhang1,Zhang2,Zhang3}, link prediction approaches
\cite{Lv1,Lv2,Lv3}, hybrid algorithms \cite{Pazzani1997,Good1999},
and so on. For a review of current progress, see Refs.
\cite{Adomavicius2005,Liu2009b} and the references therein.

\begin{figure}[ht]
\center\scalebox{0.5}[0.5]{\includegraphics{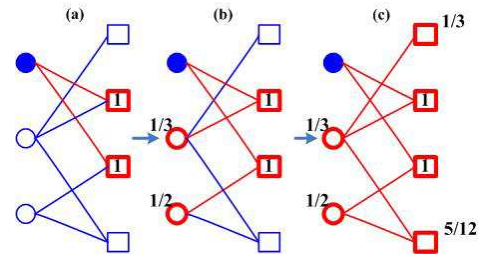}}
\caption{(Color online) Illustration of heat conduction algorithm on
a bipartite user-object network: (a) The objects collected by the
target user are activated, with temperature 1, while others are of
temperature 0. (b) Each user's temperature is the average over all
her/his collected objects. (c) Same process happens from users to
objects.}\label{Fig1}
\end{figure}

A recommender system could be described by a bipartite network
\cite{ShangEPL,LGZ2011}, in which there are two kinds of nodes:
users $U$ and objects $O$. The users' historical records are
represented by the edges connecting users and objects. Supposing
there are $m$ objects $O = \{o_1,o_2, \cdots, o_m\}$ and $n$ users
$U$ = $\{u_1, u_2,$ $\cdots,$ $u_n\}$, the system can be fully
described by an adjacency matrix ${\bf A}=\{a_{l\alpha}\}_{m,n}$,
where $a_{l\alpha}=1$ if $o_\alpha$ is collected by $u_l$, and
$a_{l\alpha}=0$ otherwise. A reasonable assumption is that objects
collected by users are what these users like and a recommendation
algorithm aims at predicting users' personal opinions on the objects
they have not yet collected \cite{Zhou2007,Zhang2007b,Zhang2007a}.
In the standard heat conduction (HC) algorithm, we first construct a
propagator matrix ${\bf W}^h$, where the element $w_{\alpha\beta}$
denotes the conduction rate from object $o_\beta$ to $o_\alpha$.
Denote ${\bf H}$ as the temperature vector of $m$ components: the
source components are of temperature 1, while the remaining
components are of temperature 0. Then the temperatures associated
with the remaining nodes could be calculated by solving the thermal
equilibrium equation ${\bf W}^h{\bf H}={\bf f}$ \cite{Zhang2007a},
where ${\bf f}$ is the flux vector. This is the discrete analog of
$-\kappa\nabla^2T(\vec{r})=\vec{\nabla}\cdot\vec{J}(\vec{r})$, where
$\kappa$ is the thermal conductivity, $\nabla^2T(\vec{r})$ is the
temperature gradient and $\vec{\nabla}\cdot\vec{J}(\vec{r})$ is the
local heat flux. In this paper, ${\bf H}(i)$ plays the role of
$-\kappa T(\vec{r})$ and ${\bf f}(i)$ plays the role of
$\vec{\nabla}\cdot\vec{J}(\vec{r})$ \cite{Zhang2007a}. In the
standard HC algorithm, the temperature of the collected objects is
constant, and the heat will diffuse from objects to users, and then
from users to objects. The temperatures of the uncollected objects
are then considered as recommendation scores: the objects given
higher temperatures would be recommended preferentially (see Fig.1
for an illustration). Since HC algorithm \cite{Zhang2007a} is
implemented based on matrix operations, it is very time-consuming
and cannot be applied to large-scale systems. Zhou {\it et
al.}\cite{PNAS} proposed a local HC algorithm, which spreads the
heat on the user-object bipartite network and can quickly generate
highly diverse yet less accurate recommendations. As a benchmark for
comparison, we call it standard HC algorithm (hereinafter, HC only
stands for local heat conduction algorithm \cite{PNAS}).

In this Brief Report, we present the biased heat conduction (BHC)
algorithm to see how objects' degrees affect the algorithmic
performance. Using data from three real systems (MovieLens, Netflix
and Delicious), we show that giving higher temperatures to the
large-degree objects than the standard HC algorithm could generate
highly accurate and diverse recommendations.


To test the performance of a recommendation algorithm, we randomly
divide a given data set into two parts: the training set and the
probe set. The information contained in the probe set is not allowed
to be used for recommendation, namely we provide a recommendation
list for each user only based on the training set. In this Brief
Report, we always keep 90\% of links in the training set and 10\% of
links in the probe set, and employ three different metrics to
measure accuracy, novelty and diversity of recommendations.

{\bf Accuracy} \cite{Zhou2007}. A good recommender algorithm should
rank preferable objects that match the user tastes in higher
positions, i.e., the objects in the probe set (indeed being
collected by users) should be put in high positions of the
recommendation list. For a user $u_i$, if the entry $u_i$-$o_j$ is
in the probe set, we measure the position of $o_j$ in the ordered
list for $u_i$. For example, if there are $100$ uncollected objects
for $u_i$ and $o_j$ is the 3rd one from the top, we say the position
of $o_j$ is $3/100$, denoted by $r_{ij}=0.03$. A good algorithm is
expected to give small $r_{ij}$. Therefore, the mean value of the
position $\langle r\rangle$ over all entries in the probe set can be
used to evaluate the algorithmic accuracy: the smaller the
\emph{average ranking score} \cite{Zhou2007}, the higher the
algorithmic accuracy.

{\bf Novelty} and {\bf diversity} \cite{Zhou2007b}. Since there are
countless channels to obtain popular objects' information,
uncovering very specific preference, corresponding to unpopular
ones, is much more significant than simply picking out what a user
likes from the list of the best sellers \cite{PNAS}. To measure this
factor, we go simultaneously in two directions: novelty (measured by
\emph{popularity}) and diversity (measured by \emph{Hamming
distance}). The popularity is defined as average degree of all
recommended objects, $\langle k\rangle$. Since it's hard for the
users to find the unpopular objects, a good algorithm should prefer
to recommend small average objects. In addition, the personalized
recommendation algorithm should present different recommendation
lists to different users according to their tastes and habits. The
diversity is quantified by the Hamming distance $S=\langle
H_{ij}\rangle$, where $H_{ij}=1-Q_{ij}(L)/L$, with $L$ is the length
of recommendation list and $Q_{ij}(L)$ is the number of overlapped
objects in $u_i$'s and $u_j$'s recommendation lists. The larger $S$
corresponds to higher diversity.

\begin{table}
\caption{Basic statistics of the tested data sets.}
\begin{center}
\begin{tabular} {ccccc}
  \hline \hline
   Data Sets      &  Users   & Objects & Links & Sparsity \\ \hline
   MovieLens      &  1,574    & 943     & 82,520 &  $5.56\times 10^{-2}$\\
   Netflix        &  10,000  & 6,000  & 701,947    & $1.17\times 10^{-2}$ \\
   Delicious      &  10,000  & 232,657 & 1,233,997 & $5.30\times 10^{-4}$\\
   \hline \hline
    \end{tabular}
\end{center}
\end{table}

Three benchmark datasets, named MovieLens, Netflix and Delicious
(See Table 1 for basic statistics), are used to test the present
algorithm. The Netflix data set is a randomly sample of huge dataset
provided for the Netflix Prize \cite{Netflix}, and the Delicious
data set is obtained by downloading publicly available data from the
social bookmarking web site Delicious.com (taking care to anonymize
user identity in the process). The Delicious data is inherently
unary while both MovieLens and Netflix data sets contain explicit
ratings from one to five. We apply a coarse-graining method to
transform them into unary forms: an object is considered to be
collected by a user only if the given rating is larger than 2. The
sparsity of the data sets is defined as the number of links divided
by the total number of user-object pairs.

\begin{table}
\caption{Algorithmic performance for {\it MovieLens, Netflix} and
{\it Delicious} data sets on the standard HC algorithm \cite{PNAS}.
The popularity $\langle k\rangle$ and diversity $S$ are obtained at
$L=10$.}
\begin{center}
\begin{tabular} {cccc}
  \hline \hline
   Data Sets     &  $\langle r\rangle$ &  $\langle k\rangle$ & $S$ \\ \hline
   MovieLens      &  0.15156 & 3.085 & 0.88196 \\
   Netflix        &  0.10629 & 1.344 & 0.86296 \\
   Delicious      &  0.26129 & 1.915 & 0.98066 \\
   \hline \hline
    \end{tabular}
\end{center}
\end{table}

Applying the standard HC algorithm on MovieLens, Netflix and
Delicious data sets, $\langle r\rangle$, $\langle k\rangle$ and $S$
are shown in Table II. One can find that although the accuracy of
the standard HC algorithm is poor, it provides highly diverse
recommendations. We argue that the less accuracy of the standard HC
algorithm lies in the fact that it assigns overwhelming priority to
the small-degree objects, leading to strong bias. Therefore, the
standard HC algorithm could be improved by reinforcing the influence
of the large-degree objects. In the last step of the standard HC
algorithm, all of the heat an object has received is divided by its
degree. Although the large-degree objects could receive lots of
heat, their temperatures are very low, while small-degree objects
would obtain high temperatures and thus be put in the top positions
of recommendation lists. A clear advantage of the standard HC
algorithm is its ability to dig out the unmainstream tastes that
almost can not be found by classical methods. However, users
generally like popular objects and thus an algorithm should also
give chance to them. We therefore propose the BHC algorithm taking
into account the object degree effect in the last diffusion step. To
an target object $o_\alpha$, instead of dividing by its degree
$k(o_{\alpha})$, the final temperature is obtained dividing by
$k^\lambda(o_{\alpha})$. The element $w_{\alpha\beta}$ of the matrix
${\bf W}^h$ would be $w_{\alpha\beta}=
\frac{1}{k^\lambda(o_{\alpha})}\sum_{l=1}^n\frac{a_{l\alpha}a_{l\beta}}{k(u_l)}$.
Comparing with the standard HC algorithm (i.e., $\lambda=1$), the
influences of large-degree objects would be strengthened if
$\lambda<1$ or depressed if $\lambda>1$.

\begin{figure}
\center\scalebox{0.33}[0.33]{\includegraphics{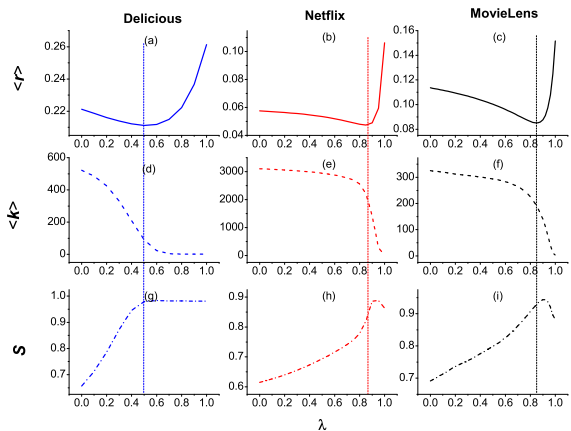}}
\caption{(Color online) Performance of the BHC algorithm on
MovieLens, Netflix and Delicious data sets. The plots (a)-(c) show
average ranking score $\langle r\rangle$ vs. $\lambda$. Subject to
$\langle r\rangle$, the optimal $\lambda_{\rm opt}$ are 0.84, 0.85
and 0.50, and the corresponding $\langle r\rangle_{\rm opt}$ are
0.0852, 0.0474, 0.2112. The plots (d)-(f) display the results for
$\langle k\rangle$ and (g)-(i) for $S$ with $L=10$. All the data
points are averaged over ten independent runs with different
divisions of training-probe sets. }\label{Fig3}
\end{figure}

\begin{figure}[ht]
\center\scalebox{0.9}[0.9]{\includegraphics{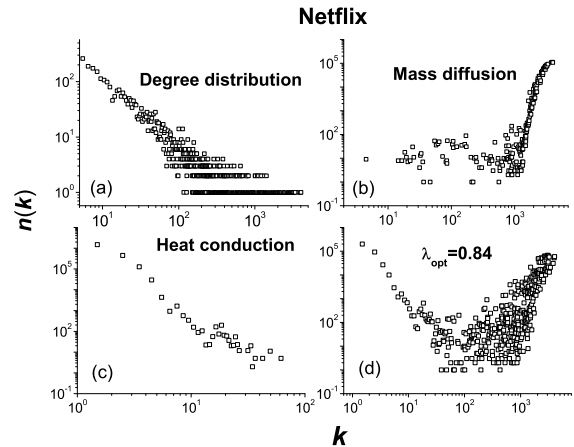}}
\caption{The plot (a) shows the object degree distribution of
Netflix data, and (b)-(d) show the correlations between the
occurrence number $n(k)$ and the object degree $k$ of MD, standard
HC and BHC algorithms when $L=10$. The results of MovieLens and
Delicious are similar.} \label{Fig5}
\end{figure}

A summary of the primary results for BHC algorithm is given in Table
III. Figure \ref{Fig3}.(a-c) report the algorithmic accuracy
$\langle r\rangle$ as a function of $\lambda$, from which one can
find that the curves obtained by BHC have clear minimums. For
example, the optimal parameter of MovieLens data is around
$\lambda_{\rm opt}=0.84$, strongly supporting our argument that the
effects of large-degree objects should be increased. Compared with
the standard case (i.e. $\lambda=1$), the average ranking score
$\langle r\rangle$ is reduced from 0.1516 to 0.0852 (improved by
43.5\%). This results indicate that giving more opportunities to the
large-degree objects will greatly increase the algorithmic accuracy.
More interestingly, when $L=10$, the Hamming distance of MovieLens
is also improved from 0.8820 to 0.9248 (see Fig. \ref{Fig3}(i)),
which is even better than 0.9173 obtained by the hybird algorithm
\cite{PNAS}. Actually, the standard HC algorithm prefers to give
more opportunities to the small-degree objects and ranks them at the
top positions of many users' recommendation lists. Therefore, the
Hamming distance may not be the highest although the popularity is
the lowest. Figure 2(b,e,h) show the similar results on Netflix,
where the optimal parameter is $\lambda_{\rm opt}=0.85$. Results of
MovieLens and Netflix are very close to each other, with the fact
that both data sets are movie-related and the sparsity is close. The
optimal parameter $\lambda_{\rm opt}$ on Delicious (See
Fig.2(a,d,g)) equals 0.5, with very small $\langle k \rangle$ and
very high $S$ ($\approx 0.98$). Both the optimal ranking score
$\langle r\rangle_{\rm opt}=0.2112$ and the Hamming distance
$S=0.9795$ of Delicious are much larger than the ones of MovieLens
and Netflix. The results are twofold: the higher sparsity of edges
and the larger number of objects. The former leads to less accurate
recommendation while the latter results in higher diversity.

\begin{table}
\caption{Algorithmic performance on BHC algorithm. The Hamming
distance is corresponding to $L=10$.}
\begin{center}
\begin{tabular} {cccccc}
  \hline \hline
   Data Sets     &  $\lambda_{\rm opt}$ &  $\langle r_{\rm opt}\rangle$ & Improvement & $S_{\rm opt}$\\ \hline
   MovieLens      &  0.84                &  0.0852                       & 43.5\%  & 0.9248 \\
   Netflix        &  0.85                &  0.0474                       & 55.4\%  & 0.8200 \\
   Delicious      &  0.50                &  0.2112                       & 19.2\%  & 0.9795 \\
   \hline \hline
    \end{tabular}
\end{center}
\end{table}

Table IV reports the performances obtained by several algorithms on
MovieLens dataset, from which one can find the accuracy $\langle
r\rangle$ of BHC algorithm is close to the result of HO-CF algorithm
which needs to compute the second-order similarity information, and
the diversity of BHC algorithm is the highest one. In order to
explain the reasons why both accuracy and diversity can be enhanced
by BHC algorithm, the frequencies of appearances $n(k)$ of objects
of degree $k$ in all users' recommendation lists are investigated.
We show the results of a typical example, Netflix, where the length
of recommendation list is $L=10$.
Different from the power-law degree distribution in
Fig.\ref{Fig5}(a), $n(k)$ of BHC algorithm has butterfly shape,
which means that the objects with large or small degrees are
recommended more frequently. Figure \ref{Fig5}(b) shows that mass
diffusion algorithm prefers to recommend the large-degree objects,
while Fig. \ref{Fig5}(c) shows that the standard HC algorithm gives
higher recommendation scores to the small-degree objects, thus the
popular objects are largely depreciated. Comparing Fig.
\ref{Fig5}(c) with Fig. \ref{Fig5}(d), at the optimal case
$\lambda_{\rm opt}=0.85$, both small-degree and large-degree objects
are recommended with high frequency by the BHC algorithm. In a word,
the advantage of BHC is that it could not only dig out the users'
very special tastes, but also find out the common interesting
objects.

\begin{table}
\caption{Algorithmic performance for \emph{MovieLens} data. $\langle
k\rangle$ and $S$ are corresponding to $L=10$. MD is abbreviations
of the algorithms proposed in Ref. \cite{Zhou2007}, Heter-NBI,
HO-CF, IMCF and WHC are abbreviations of algorithms with
heterogeneous initial resource distribution proposed in Ref.
\cite{Zhou2007b}, high-order collaborative filtering (CF) algorithm
proposed in Ref. \cite{NJP2009}, improved modified CF algorithm in
Ref. \cite{PhysicaA2010} and the algorithm presented in Ref.
\cite{LGZ2011}.} {\begin{tabular}{cccccc@{}} \toprule
  &     Algorithms     & $\langle r\rangle$  & $S$ & $\langle k\rangle$  \\
\colrule
 &  MD  & 0.1060 & 0.617 & 233  &\\
 &  HC  & 0.1516 & 0.750 & 3.09 &\\
 & Heter-NBI & 0.1010 & 0.682 & 220 & \\
 & HO-CF& 0.0826 & 0.9127& 237  & \\
 & IMCF & 0.0877 & 0.826 & 175  & \\
 & WHC  & 0.0914 & 0.941 & 179  & \\
 &  BHC & 0.0852 & 0.925 & 197  &\\ \botrule
\end{tabular}}
\end{table}

In this Brief Report, we propose a biased heat conduction algorithm
by considering the degree effects in the last step of the local heat
conduction process \cite{PNAS}, which could greatly improve the
accuracy of the standard HC algorithm. In the standard HC algorithm,
the small-degree objects are recommended overwhelmingly because in
the last step, to calculate the temperature, the received heat is
divided by the object degree. This division largely depresses the
chance of a large-degree object to be recommended. In contrast, the
power-law object degree distribution indicates that large-degree
objects are preferred by many users, therefore a good algorithm
should also pay attention to the them. In addition, a personalized
recommender system should provide each user recommendations
according to his/her own interests and habits. Therefore the
diversity of recommendation lists plays a crucial role to quantify
the personalization. The numerical results show that the
recommendation lists generated by the BHC algorithm are of
competitively higher diversity and remarkably higher accuracy than
those generated by the standard HC algorithm. The statistical
results on Facebook applications also show that the objects could be
divided into two categories \cite{JPPNAS}. One of them is collected
by almost all of users, while others are only collected by
small-size group users, which indicates that the users' tastes could
be expressed by two categories: popular one and special one.
Therefore, the reason why BHC could produce higher accuracy is that
users' two kinds of interests could be simultaneously identified.
However, how to timely track users' current popular and special
tastes is still an open problem.

We acknowledge {\it GroupLens} Research Group for providing us {\it
MovieLens} data and the Netflix Inc. for {\it Netflix} data. This
work is partially supported by the European Commission FP7 Future
and Emerging Technologies Open Scheme Project ICTeCollective
(Contract 238597), the National Natural Science Foundation of China
(Grant Nos. 10905052, and 60973069), JGL is supported by Shanghai
Leading Discipline Project (No. S30501) and Shanghai Rising-Star
Program (11QA1404500).


\begin{thebibliography}{59}

\bibitem{Broder2000}G.-Q. Zhang, G.-Q. Zhang,
Q.-F. Yang, S.-Q. Cheng, T. Zhou, New J. Phys. {\bf 10}, 123027
(2008).

\bibitem{Adomavicius2005}
G. Adomavicius, and A. Tuzhilin, IEEE Trans. Know. \& Data Eng. {\bf
17}, 734(2005).



\bibitem{Ecommerce2001}
J.~B. Schafer, J.~A. Konstan, and J. Riedl, Data Mining \& Knowledge
Discovery, {\bf 5}, 115 (2001).


\bibitem{PNAS}
T. Zhou, Z. Kuscsik, J.-G. Liu, M. Medo, J. R. Wakeling, and Y.-C.
Zhang, Proc. Natl. Acad. Sci. U.S.A. {\bf 107}, 4511 (2010).
\bibitem{Herlocker2004}
J. L. Herlocker, J. A. Konstan, K. Terveen, and J. Riedl, ACM Trans.
Inform. Syst. {\bf 22}, 5 (2004).
\bibitem{Konstan1997}
J. A. Konstan, B. N. Miller, D. Maltz, J. L. Herlocker, L. R.
Gordon, and J. Riedl, Commun. ACM {\bf 40}, 77 (1997).
\bibitem{Liu2009}
J.-G. Liu, B.-H. Wang, and Q. Guo, Int. J. Mod. Phys. C {\bf 20},
285 (2009).
\bibitem{Liu2008b}
J.-G. Liu, T. Zhou, H.-A. Che, B.-H. Wang, and Y.-C. Zhang, Physica
A {\bf 389}, 881 (2010).
\bibitem{Dun2009}
D. Sun, T. Zhou, J.-G. Liu, R. -R. Liu, C. -X. Jia, and B. -H. Wang,
Phys. Rev. E {\bf 80}, 017101 (2009).
\bibitem{Liu2009int}
J.-G. Liu, T. Zhou, B.-H. Wang, Y.-C. Zhang, and Q. Guo, Int. J.
Mod. Phys. C {\bf 21}, 137
(2009). 
\bibitem{Balab97}
 M. Balabanovi\'c and Y. Shoham, Commun. ACM {\bf 40}, 66 (1997).
\bibitem{Pazzani99} M. J. Pazzani, Artif. Intell. Rev. {\bf 13}, 393 (1999).
\bibitem{Zhang1}
M.-S. Shang, and Z.-K. Zhang, Chin. Phys. Lett. {\bf 26}, 118903
(2009).
\bibitem{Zhang2}
Z. -K. Zhang, T. Zhou, and Y.-C. Zhang, Physica A {\bf 389}, 179
(2010).
\bibitem{Zhang3}
M. -S. Shang, Z.-K. Zhang, T. Zhou, and Y.-C. Zhang, Physica A {\bf
389}, 1259 (2010).
\bibitem{Lv1}
T. Zhou, L. L\"{u}, and Y.-C. Zhang, Eur. Phys. J. B {\bf 71}, 623
(2009).
\bibitem{Lv2}
L. L\"{u} and T. Zhou, Europhys. Lett. {\bf 89}, 18001 (2010).
\bibitem{Lv3}
L. L\"{u} and T. Zhou, Physica A {\bf 390}, 1150 (2011).
\bibitem{Pazzani1997}
M. Pazzani and D. Billsus, Machine Learning {\bf 27}, 313 (1997).
\bibitem{Good1999}
N. Good, J. B. Schafer, J. A. Konstan, A. l. Borchers, B. Sarwar, J.
Herlocker, and J. Riedl,
in {\it Proceedings of the sixteenth national conference on Artificial Intellgence}, 1999, p. 439. 
\bibitem{Liu2009b}
J. -G Liu, M. Z. -Q. Chen, J. Chen, F. Deng, H. -T. Zhang, Z. -K.
Zhang, and T. Zhou. 
Int. J. Inf. Syst. Sci. {\bf 5}, 230 (2009). 
\bibitem{ShangEPL}
M.-S. Shang, L. L\"{u}, Y.-C. Zhang, and T. Zhou, Europhys. Lett.
{\bf 90}, 48006 (2010).
\bibitem{LGZ2011}
J.-G. Liu, Q. Guo, and Y.-C. Zhang, Physica A {\bf 390}, 2414
(2011).
\bibitem{Zhang2007b}Y.-C. Zhang, M. Medo, J. Ren, T. Zhou, T. Li, and F. Yang, Europhys. Lett. {\bf 80}, 68003 (2008).
\bibitem{Zhou2007} T. Zhou, J. Ren, M. Medo, and Y.-C. Zhang, Phys. Rev. E {\bf 76}, 046115 (2007).
\bibitem{Zhang2007a} Y.-C. Zhang, M. Blattner, and Y.-K. Yu, Phys.
Rev. Lett. {\bf 99}, 154301 (2007).
\bibitem{Zhou2007b} T. Zhou, L.-L. Jiang, R.-Q. Su, and
Y.-C. Zhang, Europhys. Lett. {\bf 81}, 58004 (2008).
\bibitem{NJP2009}
T. Zhou, R.-Q. Su, R.-R. Liu, L.-L. Jiang, B.-H. Wang, and Y.-C.
Zhang, New J. Phys. {\bf 11}, 123008 (2009).
\bibitem{PhysicaA2010}
J.-G. Liu, T. Zhou, H.-A. Che, B.-H. Wang, and Y.-C. Zhang, Physica
A {\bf 389}, 881 (2010).
\bibitem{Netflix}
J. Bennett and S. Lanning, 
in {\it Proceedings of the KDD Cup Workshop}, New York, 2010, p. 3.
\bibitem{JPPNAS}
J. P. Onnela and F. Reed-Tsochas, Proc. Natl. Acad. Sci. U.S.A. {\bf
107}, 18375 (2010).

\end{thebibliography}
\end{document}